\documentclass[12pt,a4paper]{article}
\usepackage{amssymb,amsfonts,amsmath}

\title{Towards better understanding of QBism}

\author{Andrei Khrennikov\\
International Center for Mathematical Modelling \\
in Physics and Cognitive Sciences\\
Linnaeus University,
V\"axj\"o, SE-351 95, Sweden\\
Andrei.Khrennikov@lnu.se}

\date{}
 
\begin{document}

\maketitle

\begin{abstract} Recently I posted  a paper entitled ``External observer reflections on QBism''. As any external observable, 
 I was not able to reflect some features of QBism properly. Therefore comments which  I received from one of its creators, C. Fuchs,   are very valuable 
- to understand better the views of QBists. Some of QBism features are very delicate and to extract them from articles of QBists 
is not a simple task. Therefore I hope that the second portion of my reflection on QBism (or better to say my reflections on Fuchs' reflections
on my reflections) might be interesting and useful for other experts in quantum foundations and quantum information theory 
(especially by taking into account  my previous aggressively anti-QBism position).  In the present 
 paper I correct some of  my previously posted critical comments on QBism. At the same time better understanding of QBists views on some 
problems leads to improvement and strengthening of other critical comments.     
\end{abstract}

\section{Introduction}

The main aim of this note  is to represent my reflections on reflections of C. Fuchs (private comments via email) 
on my recent reflections on QBism \cite{KHR_ARXIV}. However, it is useful to start with a general discussion about interpretations 
of quantum mechanics (QM)  and QBism \cite{Fuchs1}-\cite{Mermin2} as one of them. Besides such basic interpretations as the Copenhagen, nonlocal, and 
many worlds interpretations, QBism will be often compared with the V\"axj\"o interpretation  \cite{V1}, \cite{V2}. The latter is not so well popularized. However, its use as a comparative 
illustration is justified by two things: a) it was born (in 2001) as a realist answer to QBism (see \cite{Fuchs2} for QBism's reply); b) its essence (as well as in QBism) is treatment   of QM as a machinery for assignment 
of probabilities. 

We start  with the Copenhagen interpretation which  is still the basic and commonly accepted interpretation. By it QM provides 
an epistemic description of micro-phenomena, i.e., results of observations, and moreover a finer description than given by QM is impossible. The latter 
is completeness of QM  from Bohr's perspective \cite{EPR_B}.  For him \cite{BR7}, \cite{BRF} (see also  \cite{PL1}, \cite{PL3}) completeness of QM is not a 
consequence of some ``no-go theorems'' such as, e.g., von Neumann \cite{VN}
 or Bell  \cite{B} theorems, but of the existence of the indivisible quant of action given by the Planck constant $h.$ The Copenhagen interpretation is totally consistent
and  it serves quantum physics already 80 years. 

However, the essential part of the quantum community is not completely satisfied by Bohr's perspective (even those who use 
the Copenhagen interpretation as the daily routine). Theoreticians are not happy 
to recognize that their only task is to develop the operational quantum structures  which is important for applications,  but definitely boring.   Surprisingly  
even some experimenters are unhappy. (What can they want besides 
the operational interpretation of QM?)   Some of them, as, e.g., A. Zeilinger (in cooperation with C. Brukner),  are looking for some fundamental information-processing principle behind
QM \cite{Z0}, \cite{BR1} . Others (and it seems they are numerous) are not happy to study surrogates  of features of quantum systems and their own measurement devices (and the last years even their
own free wills). They want reality! They want to measure properties of quantum systems. This situation of very general dissatisfaction by the old Copenhagen perspective led to 
flowering of very exotic interpretations of QM - so exotic that the fathers of QM are spinning  in their graves. I mean nonlocal interpretations of QM\footnote{In fact, there are two 
sorts of ``quantum nonlocality''. One of them is the ontic nonlocaty - nonlocal hidden variables.  It is rooted from Bohmian mechanics. However, what is very interesting and 
not easy to explain, majority of  people saying about quantum nonlocality have in mind  the epistemic nonlocality - nonlocality of standard QM.   They say: there exists spooky action 
at a distance; this is exhibition of ``quantum nonlocality''.  The latter viewpoint has to be treated as a new interpretation of QM - the {\it  quantum nonlocality interpretation}, although people presenting 
such a nonlocal viewpoint  consider themselves as followers of the Copenhagen interpretation. However, Bohr, Heisenberg, Pauli, Landau, and Fock definitely would not support the nonlocal 
interpretation of QM.}   and the many worlds interpretation.

{\it  Quantum Bayesianism} (QBism) \cite{Fuchs1}-\cite{Mermin2}  is also considered as one of exotic interpretations,  
surprisingly as even more exotic than the two aforementioned interpretations. Why? May be because QBism is a non-realistic interpretation? However, the Copenhagen    interpretation 
is non-realistic as well... It seems that the main problem is that {\it  QBism refers to the irreducible role of a mental element in decision making about the outcomes of quantum experiments.} 
And an average modern physicist is sure that in physics  there is no place for mental elements. This position is a consequence of deep separation between cognition and psychology from one side and
physics from another. (We remind that in 19th century and at the beginning of 20th century this separation was not so deep \cite{Frontiers}.)

Can a mental element be peacefully incorporated in the body of QM? For a moment, the answer is ``may be''. On one hand, we can mention the  supporting views of Pauli (and his correspondence 
with Jung), Whitehead, and Wigner.   On the other hand, we can again point to the rise of  physical realism in the form of Bohmian mechanics, many world interpretation and the V\"axj\"o 
interpretation \cite{V1}, \cite{V2} (although all these ``realisms'' are quite exotic, respectively:  nonlocal\footnote{The main problem is not even Bohm-like nonlocality at the level of hidden variables, ontic nonlocality, 
but ``quantum nonlocality'' resulting from spooky action at a distance.}, many worlds, contextual\footnote{In the V\"axj\"o interpretation  \cite{KHR_CONT} contextuality is treated more generally 
than in modern discussions on contextuality of QM.  Context is a complex of experimental physical conditions for measurement of some observable(s), cf. with the approach developed by 
P. Grangier \cite{GR1}, \cite{GR2}.}) and attempts to exclude completely the mental element from quantum foundations.

Roughly speaking one can select between the operational, nonlocal, many worlds,  contextual and subjective interpretations of QM.  However, the presence of a mental element in decision making 
(in our case about probabilities of experimental outcomes) is essentially less mystical than, e.g., spooky action at a distance. In particular, this argument was presented by T. H\"ansch in  his talk 
at the V\"axj\"o-2015 conference in which he explained why he accepts QBism as the most natural and useful interpretation of QM, especially in the light of the quantum information revolution.

Initially my own position was characterized by the strong anti-QBísm attitude. I was not happy with the subjective interpretation \cite{Finetti} of quantum probabilities which I treated as objective 
\cite{M},  \cite{K}.
And the V\"axj\"o interpretation was created as a realist (but contextual) answer to QBism  \cite{V1}. However, QBism and the V\"axj\"o interpretation have one important common thing:
both treat QM as a formalism for prediction of probabilities. And this is the essence of the formalism, i.e., not just a supplement. The difference is that in one case probabilities are interpreted 
as subjective and in another case as objective. Both interpretations recognize the fundamental role of the Born rule. This rule is the cornerstone of QM and the rest of the quantum formalism, including 
entanglement, is just a supplement.  Moreover, both interpretations 
treat this rule as a {\it quantum generalization of the law of total probability}.\footnote{There is also a fundamental mathematical difference. The Born rule based formula of total probability (FTP) 
has very different mathematical forms which do not match each other; see \cite{Fuchs3}-\cite{Fuchs5} for the QBist version and \cite{FTPV1}-\cite{FTPV3}, \cite{KHR_CONT}  for 
the V\"axj\"o version.} 

Can one accept the use of  mental elements in physics? My viewpoint to this problem changed crucially after reading Schr\"odinger's book \cite{ST} which can be treated as an attempt of the mental 
structuring of  thermodynamics, classical and quantum. Schr\"odinger advertised the Gibbs approach to thermodynamics based on the use of virtual ensembles composed of  
{\it mental copies} of a single system.\footnote{I used this book to adapt 
the quantum formalism to applications to social science, a model of social laser \cite{Laser}.} However, in    Schr\"odinger's representation the Gibb's construction is even more subjective than in the original Gibbs writings.
Moreover, Schr\"odinger applied the same mental picture to derive quantum statistics. Hence, it seems that there is just one step to QBism?  

Schr\"odinger did not claim that the mental representation approach is identical to the real physical situation. However, he advertised this approach, because it is simpler than Boltzmann 
``physical approach'' and leads to the same answers! Why not to accept QBism by similar reason? \footnote{And I can do this without to give up the V\"axj\"o approach
which  can be considered as the quantum analog of the Boltzmann physical approach. Schr\"odinger did not say that Boltzmann was wrong ... 
Another reason for my recent movement towards QBism is my active research in applications of quantum probability in cognition, psychology \cite{Frontiers}. Here   
we need an interpretation of QM which is free from the spooky action at a distance and other quantum exotics.  I used the V\"axj\"o interpretation, but the realist contextual modeling of cognition 
is too big challenge. Therefore it is very pragmatic to use QBism \cite{CONG_QB}.  We have decision makers, they assign subjective probabilities. Moreover,  the latter are widely used in decision making 
not only  in psychological studies, but in engineering, military actions, politics, economics. Engineers  use subjective probability.  Why physicists cannot?}  

\section{My reflections on Fuchs' reflections on my reflections on QBism}

QBism is characterized  by C. Fuchs and R. Schack \cite{Fuchs5}, pp. 3-4, as follows:

{\footnotesize\em ``The fundamental primitive of QBism\index{QBism} is the concept of experience. According to QBism,
quantum mechanics is a theory that any agent can use to evaluate her expectations for
the content of her personal experience. ... An agent's beliefs and experiences are necessarily local to that agent. This implies
that the question of nonlocality simply does not arise in QBism.''} 

\subsection{QBism is not a neo-Copenhagen interpretation}

The viewpoint that QBism is a modern version of the Copenhagen interpretation is quite common. Therefore it is important to emphasize that this viewpoint is wrong.
In  \cite{KHR_ARXIV}  this problem was discussed and the position of QBists was illustrated by citation from Mermin's paper \cite{Mermin},
, p. 7-8:    

{\footnotesize\em  ``A fundamental difference between QBism and any flavor of Copenhagen, is that QBism
explicitly introduces each user of quantum mechanics into the story, together with the
world external to that user. Since every user is different, dividing the world differently
into external and internal, every application of quantum mechanics to the world must
ultimately refer, if only implicitly, to a particular user. But every version of Copenhagen
takes a view of the world that makes no reference to the particular user who is trying to
make sense of that world.''}

Thus the main difference is the private agent perspective to outcomes of experiments which is absent in the Copenhagen interpretation.
At the same time initially C. Fuchs was influenced deeply by  Pauli's version 
of the Copenhagen interpretation. However, to be consistent,  QBism cannot restrict its mental component 
to Pauli's {\footnotesize\em "objective registering apparatus, the results of which are objectively available for anyone's inspection." }  Private agent's experience is the cornerstone of 
QBism \cite{Fuchs2b}. 

\subsection{CBism or even SBism?}

In  \cite{KHR_ARXIV} it was emphasized coupling of QBism to de Finetti's subjective experience methodology of science \cite{Finetti1}. In particular, it was claimed: 
{\footnotesize\em  ``Finally we point that de Finetti was even more revolutionary than QBists, because his subjective treatment of scientific method was not restricted 
to 'special quantum world'.  …  [T]hey were not brave enough to … declare the private agent perspective for knowledge about classical world as well.''}

This claim is a consequence of my lack of education. QBists were concentrated on QM simply because it has the most severe interpretational problems which had to be solved 
to proceed successfully   towards quantum information technologies. However, they recognized from the very beginning that the power of the subjective perspective  approach 
can be used as well in classical statistical mechanics and in physics in general, moreover even outside of physics (precisely as de Finetti stated). 
In particular, in 2003 C. Fuchs pointed out (see, e.g., \cite{FuchsStruggles}, p. 812). 

{\footnotesize\em  Since becoming immersed in the subject, I have found
nothing more exciting than these trains of thought. For
they indicate the extent to which quantum foundations
research may be the tip of an iceberg—indeed, something
with the potential to drastically change our worldview,
even outside the realm of physical practice.}

Thus QBism is just a part of de Finetti's SBism (where `S' is from `Science'). Another good source on interrelation of QBism and CBism (the latter is about the private agent perspective 
for classical physics) is Mermin's paper \cite{Mermin2}. We stress that in this paper the mental dimension is especially strong.  

\subsection{Born's rule as generalization of the formula of total probability}

As was mentioned in introduction, QBists treat Born's rule as quantum generalization of  FTP. We remind that classical FTP functionally  connects the probability distribution of 
one observable, say $A,$ 
with probability distribution of another observable, say $B,$ by using the conditional probabilities $p(B\vert A):$ 
\begin{equation}
\label{FTP}
p(b_j)= f(p(a_i), p(b_j\vert a_i)),
\end{equation}
where 
\begin{equation}
\label{FTP1}
f(x,y)  =\sum_i x_i y_j.
\end{equation}
By considering two quantum observables, e.g., given by POVMs, we 
can represent the Born rule as a similar transformation. It is important to remark that even in classical case we can find a variety   of functional representations (\ref{FTP}). The form (\ref{FTP1}) of 
the function $f$ is in some sense the best adapted to the Bayes formula for classical conditional probability. Hence, in the quantum case we also can construct a variety of the functional 
representations of the form (\ref{FTP}). QBists selected one special form of $f$ coupled to so called SIC-POVMs.
 
 In  \cite{KHR_ARXIV} I was very critical to ``addiction'' of QBists to such special POVMs.  From my viewpoint, it would be natural at least to start with an arbitrary informationally compelte POVM.
{\footnotesize\em   ``However, for QBists the above generalization – to start the probability update scheme with an arbitrary POVM measurement and not with the SIC-POVM – 
seems to be unacceptable.  They are really addicted on SIC-POVMs and on completeness of information gained at the first step, information about the state, even 
at the price of appearance of counterfactuals,''}  \cite{KHR_ARXIV}.  In his comments C. Fuchs motivates the use of the SIC-POVM FTP representation of  the Born rule   as  encoding 
{\footnotesize\em as much unique Hilbert space structure  into the Born rule as possible.} It seems that QBists hope to obtain QM from this  SIC-POVM FTP. 
And this is an exciting, but very challenging project! 

The {\it SIC-POVM FTP is the basic axiom of QBism.} What is it probabilistic meaning?

In  \cite{KHR_ARXIV} I interpreted SIC-POVM FTP as the quantum rule for probability update (PU). This matches  generalized classical PU known as {\it Jeffrey's conditioning}
\cite{j1}, \cite{j2}.
This my viewpoint   on SIC-POVM FTP resulted from my misunderstanding of the basic principle of QBism and exploring the analogy with the V\"axj\"o interpretation,
where the quantum FTP (so to say, two arbitrary POVMs FTP) is interpreted as the quantum analog of Jeffrey's  PU. 
This is the delicate issue and we start the discussion from reminding the classical PU and its generalization - Jeffrey's  PU. 

\subsection{Classical Bayesian probability update}

 The probability of a hypothesis $H$ conditional on a collected data $E$  is given  
{\it Bayes' formula} - the definition of conditional probability:
\begin{equation}
\label{BF} 
p(H\vert E) = p(H \& E)/p(E),  p(E) > 0.
\end{equation}
{\it Bayes' Theorem} relates the ``direct" probability of a hypothesis 
conditional on the data, $p(H\vert E),$ to the ``inverse" probability of the data conditional on the hypothesis, $p(E\vert H).$
\begin{equation}
\label{BF1} 
p(H\vert E) = [p(H)/p(E)] p(E\vert H), 
\end{equation}   	
This possibility to ``invert'' probability $p(H\vert E),$  is based on commutativity of the operation of conjunction in  Boolean logic:
$H \& E= E \& H.$ Of course, the definition of conditional probability by Bayes' formula is possible only under this condition. 

Subjectivists think of learning as a process of belief revision in which a prior subjective probability $p$ is replaced by a posterior probability 
$q$ that incorporates newly acquired information. This process proceeds in two stages. First, some of the subject's probabilities are directly altered by experience, 
intuition, memory, or some other non-inferential learning process. Second, the subject ``updates" the rest of her opinions to bring them into line with 
her newly acquired knowledge.

{{\bf  	Simple Conditioning.} If a person with a prior such that $0 < p(E) < 1$ has a learning experience whose sole immediate 
effect is to raise her subjective probability for $E$ to 1, then her post-learning posterior for any proposition $H$ should be 
\begin{equation}
\label{BF27} 
p(H) = p(H\vert E).
\end{equation}

Though useful as an ideal, simple conditioning is not widely applicable because it requires the learner to become absolutely certain of $E$'s truth. 
As R. Jeffrey has argued \cite{j1}, \cite{j2} the evidence we receive is often sufficient only to assign some probabilities to occurrence of $E.$ 
Here the direct effect of a learning experience will be to alter the subjective probability of some proposition 
without raising it to 1 or lowering it to 0. Experiences of this sort are appropriately modeled by what has come to be called {\it Jeffrey conditioning.}

{\bf Jeffrey Conditioning.}
If a person  has a learning experience whose sole immediate effect is to change 
her subjective probability for $E$ to $p(E)$, then her post-learning posterior for any $H$ should be given by FTP: 
\begin{equation}
\label{LFTP}
p(H) = p(E) p(H\vert E) + (1 -  p(E))p(H\vert \overline{E}),
\end{equation}
where, for any proposition $F,$ the symbol $\overline{F}$ denotes negation of the proposition $F.$
Obviously, Jeffrey conditioning reduces to simple conditioning when $p(E) = 1.$
Jeffrey conditioning can be generalized to the case of a collection of hypotheses and pieces of data represented mathematically as the disjoint partitions of the space of elementary events $\Omega,$ 
$(H_j)$ and $(E_i),$ where  $H_i   \& H_j=\emptyset, E_i   \& E_j=\emptyset, i \not= j:$ 
 \begin{equation}
\label{LFTP3}
p(H_j) = \sum_i p(E_i) p(H_j\vert E_i).
\end{equation}
For further considerations, it is useful to represent Jeffrey's PU in terms of two observables, $A$ and $B:$ the events $E_i$ correspond to observations of the values 
$a_i$ of $A$ and the hypotheses $H_j$ are about (possible) observations of the values $b_j$ of $B.$ Then the PU rule (\ref{LFTP3}) coincides with FTP   (\ref{FTP}),
(\ref{FTP1}). 

In the pure subjective probability approach all probabilities in the right-hand side of  (\ref{LFTP3}) are treated as subjective. However, in classical PU it is quite common to use 
mixed subjective-objective PU. Here the probabilities $p(E_i)$ are considered as subjective, but the conditional probabilities    $p(H_j\vert E_i)$ as objective. (In the observational notations 
they have the form $p(b_j\vert a_i).)$  The latter probabilities
were collected, e.g., by using frequencies of observations of the hypotheses $H_j$ on the basis of events $E_i.$ These are ``structural constants'' of the update.
Soon we shall  discuss these interpretations of  PU given by (\ref{LFTP3})  in the quantum framework.

\subsection{SIC-POVM formula of total probability and probability update} 

Jeffrey's PU can practically automatically generalized to the quantum case by using POVMs representation of observables $A$ and $B$ and the 
definition of quantum conditional probability. There is a quantum state $\rho.$ The information about it is updated as the result of $A$-measurement, i.e.,  in general
we do not try to reconstruct $\rho$ completely, but we are fine by knowing just  the information gained from the $A$-measurement. (Of course, sometimes we can be lucky and be able 
to perform information-complete measurement.)   On the basis of this 
information and by knowing probabilities gained from sequential measurements, first $A$ and then $B,$ we make PU of probabilities for the possible values 
of the $B$-observable.  In this way a quantum analog of FTP is interpreted in the V\"axj\"o framework \cite{FTPV1}-\cite{FTPV3}, \cite{KHR_CONT}. 
Here all probabilities are interpreted objectively.
However, there is nothing wrong to proceed with subjective probabilities, especially in the framework of subjective-objective PU. In the very important case 
of the observable $A$ of von Neumann-L\"uders type having nondegenerate spectrum, the conditional (or better to say transition) probabilities    $p(b_j\vert a_i)$
do not depend on the state $\rho.$ It is natural to consider them as objectively determined  ``structural constants". 

In \cite{KHR_ARXIV} this viewpoint on quantum versions of FTP, as generalized PU,  was extrapolated to QBism with subjective interpretation of probabilities as  its main specialty,
as well as the special SIC-POVM form of FTP:
\begin{equation}
\label{SIC}
p(b_j) =  \sum_i \Big( (d+1) p(a_i) - \frac{1}{d}\Big) p(b_j\vert a_i),
\end{equation}
where $d$ is the dimension of the state space.

However, it happens that such extrapolation was not justified and this is real misunderstanding of QBism's interpretation of the quantum FTP (in their SIC-POVM form).
QBists do not consider   SIC-POVM FTP as a generalization of classical Jeffrey's PU . Their interpretation is more delicate \cite{Fuchs3a}. Here I prefer to cite C. Fuchs (email correspondence):
{\footnotesize\em  ``I understand what the FTP is, but I would never call it
a machine for "updating" probabilities.  When I think of  ``updating
probabilities'' the kind of apparatus that comes to mind is, for instance,
an application of Bayes rule for conditionalizing, or Jeffrey's
conditionalization rule, etc.  That is,  ``updating'' is about changing
probabilities upon the acquisition of new information.  But the FTP is not
about updating in that sense.  Rather QBism views it as a ``coherence''
statement in the sense of de Finetti's Dutch book argument. That is, it is a
relation between probability assignments, all defined at the same time
(i.e., synchronically).  It is not about the ``changing of probabilities,''
but about how various assignments should fit together to start with.
Similarly, this is how we think of the Born Rule when viewed as a modified
FTP:  It as a specification for how various probabilities defined at one
time should fit together.  It is not about the changing of probabilities
when information is acquired.''}
We also present  another of Fuchs' communications clarifying this viewpoint:

{\footnotesize\em  [W]e think of it [the SIC-POVM FTP] as a synchronic
coherence requirement (much like a Dutch book argument for the FTP, which is
purely synchronic).  Its role is to say that an agent should NOT let his
probability assignment for the outcomes of a given experiment fly free from
the assignments he would make in a hypothetical (or counterfactual)
experiment involving an intermediate SIC.  The assignments should be related
even though not both experiments can be performed at the same time.}   See
also section 4  \cite{Fuchs5} for detail.

The aforementioned synchronization of probabilities is purely subjective; even conditional probabilities  (as the quantum state
assignment), {\it else QBism would be inconsistent.} It is important to remark that QBism as any interpretation of QM evolves 
and have some flavors\footnote{This is not the feature of only 
QBism. The same can be said about, e.g., the Copenhagen interpretation, A. Plotnitsky even invented the terminology ``interpretation in the spirit of Copenhagen'' \cite{PL1}, \cite{PL3}.}; 
in the ``old QBism'' of Caves, Fuchs, and  Schack the state and the conditional probabilities were treated objectively!
We can refer to  sections 7 and 8 of \cite{Fuchs1}.

\subsection{Quantum Bayesian agents}

In  \cite{KHR_ARXIV} it was pointed out ,  {\footnotesize\em ``It is important that QBism uses this rule as an information constraint to determine a class of so to say `quantum agents', i.e., those 
who `get tickets to the QBism performance.'  Thus private users of QM are those who know the main rule of the game."} And this matches completely QBism. 
The idea is that QM is something used only by a privileged class of people.  Those educated in the methods of QM are able to make 
better decisions (because of certain basic features of nature) than those not educated in the methods of QM.  

However, the attempt  \cite{KHR_ARXIV} to introduce into QBism,  so to say, 
the ``universal quantum Bayesian agent'', in the spirit of Brukner \cite{BR4} (who wrote about the ``hypothetical agent''), are not welcome in 
QBism - experience has to be really private. 

\section{Understanding QBism}

We now summarize our discussion on distinguishing and delicate features of QBism:

\begin{enumerate}
 \item QBism is about private experience of agents making predictions about outcomes of experiments. The sample of agents is not arbitrary. A QBism agent has to belong to so to say 
``quantum club'', i.e., to be ``quantumly educated.'' 
\item  The paradigm of the ``universal quantum Bayesian agent'' is totally foreign to QBism. 
\item  It is the natural further step from exploring mental structures in statistical physics and thermodynamics - technique of calculation of probabilities based on 
invention of virtual ensembles (Gibbs, Schr\"odinger, Jaynes).  The use of such ensembles composed of mental copies of a single system naturally leads to 
the subjective interpretation of probabilities. 
\item It matches well with scientific methodology presented by de Finettei in his great pamphlet ``Probabilismo'' \cite{Finetti1}. 
 By this methodology science is about our private experiences.
QBists understand well that QBism is a part of so to say SBism, where ``S'' is for science. In particular, CBism (where ``C'' from classical physics) was 
discussed in very details by D. Mermin. S. Fuchs started his pathway to QBism from subjective treatment of classical thermodynamics in the spirit of 1).  
Concentration on QBism is explained by real necessity to solve 
the interpretational problems of QM in the light of the quantum information revolution. 
\item  QBism is a local interpretation of QM.
\item Is QBism a non-realist interpretation of QM? It is a complicated philosophic issue. It seems that for subjectivists (both classical, as de Finetti, and quantum) the reality 
is constructed from our private experiences. From this viewpoint they are realists.
\item  QBism is not a version of Copenhagen interpretation, although it was rooted in it.
\item  By QBism the quantum formalism is a machinery for consistent assignment of subjective probabilities for outputs of possible experiment. 
\item  This probability synchronization machinery cannot be simply treated as a kind of probability update machinery.\footnote{This is the main difference between QBism and the V\"axj\"o interpretation of QM 
(more fundamental than the difference in the interpretation of probabilities).}
\item The Born rule is treated as the main axiom of QM, other axioms are just supplement.
\item This rule is represented in the form of generalized law of total probability SIC POVM FTP.\footnote{This quantum analog of FTP differs crucially from the V\"axj\"o version of 
quantum FTP (which is an additive perturbation of classical FTP).} 
\item The use of SIC POVMs is crucial, since QBists hope that SIC POVM FTP encodes all basic features of QM. 
\item For consistency of QBism all probabilities in SIC POVM FTP have to be interpreted as subjective probabilities, even conditional probabilities $p(b_j\vert a_i).$\footnote{Since SIC POVMs 
are based on one dimensional projectors, the corresponding conditional probabilities do not depend the initial quantum state $\rho.$ Hence, it is very attractive to treat them as objectively 
determined (as features of observables) structure constants. However, ``modern QBists'' would not accept such an interpretation.}
\end{enumerate}

\section*{Acknowledgments} 

I would like to thank C. Fuchs for his detailed comments on my paper \cite{KHR_ARXIV} and for discussions about QBism and interpretations of quantum probabilities 
(during the V\"axj\"o series of conferences on quantum foundations,  2001-15).


\begin{thebibliography}{199}

\bibitem{KHR_ARXIV} A. Khrennikov, External observer reflections on QBism. 	arXiv:1512.07195 [quant-ph]

\bibitem{Fuchs1}  Fuchs, C.~A. (2002). Quantum mechanics as quantum information (and only a little more), in A. Khrennikov (ed.), \emph{Quantum Theory: Reconsideration
of Foundations, Ser. Math. Modeling} {\bf 2} (V\"axj\"o University Press, V\"axj\"o), pp. 463--543. 

\bibitem{Fuchs2} Fuchs, C.~A. (2002). The anti-V\"axj\"o interpretation of quantum mechanics, \emph{Quantum Theory: Reconsideration of Foundations, 
Ser. Math. Model.} {\bf 2} (V\"axj\"o University Press, V\"axj\"o), pp. 99--116. 

\bibitem{Caves1}  Caves, C.~M., Fuchs, C.~A. and Schack, R. (2002). Quantum probabilities as Bayesian probabilities, \emph{Phys. Rev. A} {\bf 65}, 022305.

\bibitem{Caves2}  Caves, C.~M., Fuchs, C.~A. and Schack, R. (2007). Subjective Probability and Quantum Certainty, \emph{Stud. Hist. Phil. Mod. Phys.} {\bf 38}, 255.

\bibitem{Fuchs2a} Fuchs, C. (2007). Delirium quantum (or, where I will take quantum mechanics if it will let me), in G. Adenier, C. Fuchs and A.~Yu. Khrennikov (eds.), 
Foundations of Probability and Physics-3, \emph{Ser. Conference Proceedings} {\bf 889} (American Institute of Physics,  Melville, NY), pp. 438--462.  

\bibitem{Fuchs2b} Fuchs, C. (2012). QBism, the Perimeter of Quantum Bayesianism. 	arXiv:1003.5209 [quant-ph].

\bibitem{Fuchs3} Fuchs, C.~A. and Schack, R. (2011). A Quantum-Bayesian Route to Quantum-State Space, \emph{Found. Phys.} {\bf 41}, p. 345.

\bibitem{Fuchs3a} Fuchs, C.~A. and Schack, R. (2012).  Bayesian Conditioning, the Reflection Principle, and Quantum Decoherence.
In: Probability in Physics, ser. The Frontiers Collection,  Springer, Berlin-Heidelberg, pp. 233-247

\bibitem{Fuchs4} Fuchs, C.~A. and Schack, R. (2013). Quantum-Bayesian Coherence, \emph{Rev. Mod. Phys.} {\bf 85}, p. 1693.

\bibitem{Fuchs5} Fuchs, C.~A. and Schack, R. (2014). QBism and the Greeks: why a quantum state does not represent an element of physical reality,    \emph{Phys. Scr.}  {\bf 90},  015104. 

\bibitem{Fuchs6} Fuchs, C.~A.,  Mermin, N.~D. and Schack, R. (2014). An Introduction to QBism with an Application to the Locality of Quantum Mechanics, \emph{Am. J. Phys.} {\bf 82}, p. 749.

\bibitem{FuchsStruggles} C. A. Fuchs, My Struggles with the Block Universe, (2014),  available at http://arxiv.org/abs/1405.2390.

\bibitem{Mermin}   Mermin, N.~D.  (2014).  Why QBism is not the Copenhagen interpretation and what John Bell might have thought of it, 
Preprint: arXiv:1409.2454v1 [quant-ph].

\bibitem{Mermin2} D, Mermin (2014), QBism put the scientist back to science.  \emph{Nature} {\bf 507}, N 7493, 421-423.

\bibitem{V1}  Khrennikov, A. (2002). V\"axj\"o interpretation of quantum mechanics, in   \emph{Quantum Theory: Reconsideration of Foundations.} 
(V\"axj\"o Univ. Press), pp. 163--170; Preprint:  arXiv:quant-ph/0202107.

\bibitem{V2}  Khrennikov, A. (2004). V\"axj\"o interpretation-2003: Realism of contexts, in \emph{Proc. Int. Conf. Quantum Theory: Reconsideration of
Foundations, Ser. Math. Modelling in Phys., Engin., and Cogn. Sc.} {\bf 10} (V\"axj\"o Univ. Press), pp. 323--338.

\bibitem{EPR_B} Bohr, N. (1935). Can quantum-mechanical description of physical reality be considered complete?  \emph{Phys. Rev.} {\bf 48}, pp. 696--702.

\bibitem{BR7}  Bohr, N. (1938). The Causality Problem in Atomic Physics, in  J. Faye and H.~J. Folse (eds.), (1987). \emph{The Philosophical Writings of Niels Bohr, Volume 4: Causality  and Complementarity, Supplementary Papers} (Ox Bow Press; Woodbridge, CT), pp. 94--121.  

\bibitem{BRF}  Bohr, N. (1987). \emph{The Philosophical Writings of Niels Bohr}, 3 vols. (Ox Bow Press, Woodbridge, CT).

\bibitem{PL1} Plotnitsky, A. (2006). \emph{Reading Bohr: Physics and Philosophy} (Springer, Heidelberg-Berlin-New York). 

\bibitem{PL3}  Plotnitsky, A. (2012). \emph{Niels Bohr and Complementarity: An Introduction} (Springer, Berlin and New York).

\bibitem{VN} von Neumann, J. (1955). \emph{Mathematical foundations of quantum mechanics} (Princeton Univ. Press, Princenton).

\bibitem{B} Bell, J.: Speakable and Unspeakable in Quantum Mechanics. Cambridge Univ. Press, Cambridge (1987)

\bibitem{Z0} Zeilinger, A.  (1999). A foundational principle for quantum mechanics, \emph{Foundations of Physics} {\bf 29}, 6pp. 31--641.

\bibitem{BR1}    Brukner, C.  and Zeilinger, A. (1999). Malus' law and quantum information, \emph{Acta Physica Slovava} {\bf 49} 4, pp. 647--652.

\bibitem{Frontiers} A. Khrennikov, Quantum-like modeling of cognition. {\it Frontiers in Physics}, {\bf 3}, art. 77 (2015).

\bibitem{KHR_CONT} Khrennikov,  A. ( 2009). {\it Contextual Approach to Quantum Formalism,} (Springer, Berlin-Heidelberg-New York)

\bibitem{GR1} P. Grangier, Contextual objectivity: a realistic interpretation of quantum mechanics, European Journal of Physics 23, 331 (2002)  [arXiv:quant-ph/0012122]. 

\bibitem{GR2} P. Grangier, Contextual objectivity and the quantum formalism, IJQI,  3 (1): 17-22 (2005) [arXiv:quant-ph/0407025]

\bibitem{Finetti} De Finetti, B. (2008). \emph{Philosophical Lectures on Probability} (Springer Verlag, Berlin-New York). 

\bibitem{M} von Mises, R. (1957). \emph{Probability, Statistics and Truth} (Macmillan, London).

\bibitem{K} Kolmolgoroff, A.~N.  (1933). \emph{Grundbegriffe der Wahrscheinlichkeitsrechnung} (Springer-Verlag, Berlin).

\bibitem{FTPV1}  Khrennikov,  A.~Yu. (1999). \emph{ Interpretations of Probability} (VSP Int. Sc. Publishers, Utrecht/Tokyo).

\bibitem{FTPV2} Khrennikov,  A.~Yu. (2001). Linear representations of probabilistic transformations  induced by context transitions, \emph{ J. Phys.A: Math. Gen.} {\bf 34}, pp. 9965--9981.

\bibitem{FTPV3} Khrennikov,  A.~Yu. (2001). Origin of quantum probabilities, in A. Khrennikov (ed.), \emph{ Foundations of Probability and Physics} (V\"axj\"o-2000, Sweden; WSP, Singapore), pp. 180--200.

\bibitem{ST} Schr\"odinger, E. 1989 Statistical thermodynamics. Dover Publications.

\bibitem{Laser} A. Khrennikov, 2015 Towards information lasers. \emph{ Entropy},  {\bf 17}, N 10, 6969-6994.

\bibitem{CONG_QB} E. Haven and A. Khrennikov 2015, Statistical and subjective interpretations of probability in quantum-like models of cognition and decision making.
{\it J. Math. Psych.} http://www.sciencedirect.com/science/article/pii/S0022249616000213

\bibitem{Finetti1}  De Finetti, B. (1931). Probabilismo, \emph{Logos} {\bf 14}, p. 163; transl. (1989). Probabilism, \emph{Erkenntnis} {\bf 31}, pp. 169--223.

\bibitem{j1} Jeffrey R. 1987 Alias Smith and Jones: The testimony of the senses, \textit{Erkenntnis} \textbf{26}, 391-399.

\bibitem{j2} Jeffrey R.  1992 \textit{Probability and the art of judgment.} New York: Cambridge University Press.

\bibitem{BR4} Brukner, C.,   \emph{On the quantum measurement problem}, Preprint	arXiv:1507.05255 [quant-ph]. To be published  in 
Proc.  Conf. {\it Quantum UnSpeakables II: 50 Years of Bell's Theorem} (Vienna, 19-22 June 2014).

  
\end{thebibliography}
\end{document}